# Simulation of laser induced retinal thermal injuries for non-uniform irradiance profiles and their evaluation according to the laser safety standard

Kotzur, Sebastian, Wahl, Siegfried, Frederiksen, Annette

# Simulation of laser induced retinal thermal injuries for non-uniform irradiance profiles and their evaluation according to the laser safety standard

Sebastian Kotzur[a,b], Siegfried Wahl[b,c], and Annette Frederiksen[d]

[a]Robert Bosch GmbH, Corporate Research, 71272 Renningen, Germany
[b]Institute for Ophthalmic Research, Eberhard Karls University Tübingen, 72076 Tübingen, Germany
[c]Carl ZEISS Vision International GmbH, 73430 Aalen, Germany
[d]Robert Bosch GmbH, Chassis Systems Control, 71701 Schwieberdingen, Germany

**ABSTRACT**

Laser systems emitting radiation in the visible and near infrared region are potentially hazardous for the retina of the human eye. This can result in irreparable injuries due to photomechanical, photothermal or photochemical light-tissue interactions. This investigation focuses on the photothermal interaction for which a computer model is used to simulate the thermal behavior of the retina and to predict the injury threshold values. The most important factors are the wavelength of the radiation, the exposure time and the irradiance profile on the retina. For performing safety evaluations and classifications the laser safety standard IEC 60825-1:2014 has to be considered. These evaluations are based on emission limits which depend on the same above mentioned factors. According to the IEC 60825-1:2014, non-uniform retinal images are treated by an image analysis where an averaged spot size is used. This averaged size is calculated by the extent of the irradiance profile along two orthogonal directions. Unlike the laser safety standard, the computer model predicts the injury thresholds for an irradiance profile on the retina without averaging the spot size. In this investigation, a broad variety of non-uniform retinal images is investigated with regard to the injury thresholds predicted by the computer model and to the classifications according to the laser safety standard.

## 1. INTRODUCTION

A laser represents a potential hazard for the human eye and the skin. Depending on the wavelength and pulse duration of the laser radiation different damage mechanism can lead to injuries. The ICNIRP guidelines[1] provide exposure limits ensuring protection against all possible injuries. Another common document which provides corresponding emission limits is the laser safety standard IEC 60825-1:2014.[2] This standards is a product safety standard and is used for the classification of laser systems. Here, a laser system of Class 1 is considered as a safe product.

This investigation focuses on the retinal thermal injury where a denaturation process can irreversibly damage the retina leading to a permanent vision loss. The retinal thermal injury is induced by radiation in the visible and near-infrared region and for pulse durations in the microsecond and second regime. The corresponding emission limits are derived from experimental threshold values.[3] A lot of measurements were performed with non-human primates (NHP) which represent a suitable model to derive emission limits as their retinas are more sensitive compared to the human retina.[4–6] The measured threshold values are often referred to as $ED_{50}$ values since there is a probability of 50% to induce an injury. In the scope of this investigation the term threshold (THR) value will be used for the $ED_{50}$ value. The emission limits are obtained by applying a safety factor, here referred to as reduction factor (RF), to the threshold values.

However, all threshold experiments which are used to derive the emission limits are based on investigations of circular symmetric laser radiation with a Gaussian or tophat shape. For the eye safety analysis of arbitrary non-uniform retinal irradiance distribution the laser safety standard[2] and the ICNIRP guidelines[1] instruct to measure the amount of the retinal radiant energy which is contained in a field stop and to compare it to the limit. To be more precise, the field stop has to be varied in its size, position and orientation to obtain the most restrictive result.[7] It is stated that the averaged field stop size and therefore the averaged retinal spot size is used for the calculation of the emission limit without giving a rationale. As there is currently no experimental data available, the question arises whether this procedure is justified or not.

The main focus of this investigation is the consideration of non-uniform spots in the thermal damage regime with regard to the threshold value and the emission limit according to the laser safety standard IEC 60825-1:2014.[2] The threshold values are obtained by a computer model. The model is validated against retinal thermal injury $ED_{50}$ data which are based on circular symmetric irradiance profiles. The computer model predicts the threshold value and is valid for pulse durations longer than 100 $\mu$s. The topic of non-uniform irradiance profiles concerns for example line lasers which have a collimated and a divergent axis. These systems are generally allowed to have a higher output power than laser pointers as the radiation is imaged on a larger area on the retina. However, the retinal image can be basically described by a thin long line radiating the retina. A classification according to the laser safety standard IEC 60825-1[2] is based on the averaged spot size.

## 2. RETINAL THERMAL INJURY MECHANISM

The retinal thermal injury mechanism is dominant for exposure durations between the microsecond and second regime. It is a denaturation of the proteins within the retina and can be described as a thermal burn. As a consequence, necrosis of the cells occurs leading to a permanent vision loss in the concerned regions. If the retina is exposed to laser radiation, most of the radiant energy will be absorbed by the retinal pigment epithelium (RPE).[8,9] The energy is converted into thermal energy. In the RPE, the temperature increase is the highest.

The injury can be described analytically with the Arrhenius equation[10,11]

$$\Omega(t) = A \int_0^t \exp\left(-\frac{E}{RT(t')}\right) dt'. \tag{1}$$

Here, $R$ is the ideal gas constant. $A$ is a frequency factor and $E$ is the inactivation energy. Both of these values can be measured experimentally.[10] The temperature dependence of $A$ is negligible small compared to the exponential function in the integrand and is assumed to be constant.[12] The value $\Omega$ represents the percentage of denatured molecules with the relation

$$c_{\text{denatured}}(t) = 1 - \exp(-\Omega(t)). \tag{2}$$

In general, a thermal injury occurs for $\Omega > 1$. The computer model in this investigation is based on the definition of a damage threshold where $\Omega$ equals one. This corresponds to a degree of denaturation of 63% which is commonly accepted to define the thermal injury.[13]

Equation (1) shows that the criterion for the denaturation process is not based on a temperature limit but rather on the temporal temperature profile. For this reason, the thermal damage is strongly coupled to the shape of the retinal image and the exposure duration.

## 3. COMPUTER MODEL TO DESCRIBE LASER INDUCED RETINAL THERMAL INJURIES

The computer model used in this investigation is a further development of the Seibersdorf Laboratories Model (SLM). The SLM was build by Jean who validated his model against 31 studies with 253 experimental $ED_{50}$ values.[13,14] The SLM is a two dimensional axial symmetric model for investigating circular symmetric irradiance profiles. The model predicts the $ED_{50}$ threshold value in terms of the total intraocular energy of the retinal irradiance profile. The principal procedure to build a retinal thermal injury model is to define a heat source within the retinal layer with the irradiance profile. Then, the heat transfer equation is numerically solved and the temporal temperature behavior is used in the Arrhenius equation (1).[15–17]

In the following, the main elements of the SLM are presented which are based on the validation against non-human primates. For obtaining the retinal image size, the LeGrand model for the relaxed human eye was used.[18] Here, the eye model consists of four layers, namely the cornea, the aqueous, the lens and the vitreous body. The sizes of these layers were scaled with a factor of 0.8 to approximately obtain the eye properties of the non-human primate.[8,19] As laser radiation cannot be focussed on an infinitely small spot, there is a minimum spot size for the retinal image. The measurements on non-human primates show that the threshold values converge to a constant energy for laser sources smaller than 5 mrad.[20–23] This is set as the minimum spot size in the computer model. The propagation of the laser radiation through the eye is accompanied by absorption at the layers of the eye model and by intraocular scattering.[24,25] The amount of radiant energy that reaches the retina depends on the wavelength and on the beam size. It is distinguished between the total transmission $T_{\text{Total}}$ which applies for relatively large retinal images and the direct transmission $T_{\text{Direct}}$[26] which applies for small retinal images.

It is suggested by Jean[14] to use an equation that ensures a continuous transition from small to large retinal spot sizes. By describing the retinal spot size with the radius $r_{\text{Spot}}$ the effective transmission is given by

$$T_{\text{eff}}(\lambda, r_{\text{Spot}}) = T_{\text{Total}}(1 - g(\lambda) h(r_{\text{Spot}})) \quad (3)$$

with

$$g(\lambda) = \frac{1}{2} \exp\left(-\frac{\lambda}{883\,\text{nm}}\right) \quad (4)$$

and

$$h(r) = \exp\left(-\frac{2r}{600\,\mu\text{m}}\right). \quad (5)$$

The amount of radiant energy that is absorbed by the retinal layers will be directly converted into thermal energy and therefore the absorption behavior of the tissue defines a heat source $q(t, \vec{r})$. The heat transfer equation is given by

$$\rho C \frac{\partial T}{\partial t} = k\Delta T + q(t, \vec{r}). \quad (6)$$

The thermal properties of the tissue are listed in Table 1. They are assumed to be homogenous and similar to water since more than 80% of the retina consists of water.[27] Furthermore, the thermal properties are assumed to be independent of the temperature which results in a linear behavior with regard to the heat source or in case of a laser irradiated retina, to the laser power.

Table 1. Thermal properties of the retina to simulate the heat transfer by a laser source.[14]

| thermal property | value |
|---|---|
| conductivity $k$ | $0.6305\,\frac{\text{W}}{\text{m K}}$ |
| specific heat $C$ | $4178\,\frac{\text{J}}{\text{kg K}}$ |
| density $\rho$ | $992\,\frac{\text{kg}}{\text{m}^3}$ |
| initial temperature $T_0$ | $310.5\,\text{K}$ |

To determine the thermal injury threshold value, the Arrhenius equation (1) has to be solved. Here, the temporal temperature behavior of the RPE layer is needed since the lesion is build within this layer.[8,9,28] In addition, the model was validated with the definition that there is MVL which has a diameter of 50 µm for the non-human primate. Therefore, to determine the threshold value the temperature in the RPE layer at the edge of the MVL circle has to be considered. This will result in a value for $\Omega \geq 1$ within the MVL area where all enclosed cells are damaged.[29]

To obtain the threshold values for human eyes, the following adjustments of the above described model have to be made. The minimum visible lesion is set to 20 µm even though such small lesions are not detected by ophthalmoscopes.[30] Furthermore, the minimum spot size is set to 25 µm.[1,2,29,31] At last, the focal length of the relaxed human eye is set to 16.68 mm which correspond to an air-equivalent focal length of the Le Grand eye model.[18,32]

# 4. THREE DIMENSIONAL COMPUTER MODEL TO DESCRIBE THERMAL INJURIES OF NON-UNIFORM IRRADIANCE PROFILES

We improved the SLM and developed a three dimensional model to investigate non-uniform retinal images. In this investigation two different irradiance profiles are considered, the elliptical tophat and the rectangular tophat. Both of the irradiance profiles are illustrated in Fig. 1. As both profiles are axisymmetric the heat transfer equation (6) can be simplified by defining two symmetry planes where only two octants of the three-dimensional model have to be solved. In the following, the irradiance profiles are given in a coordinate system with an $x$- and $y$-axis. Both profiles are characterized by the two diameters $d_x$ and $d_y$ representing the extent of the image in the corresponding orthogonal axis.

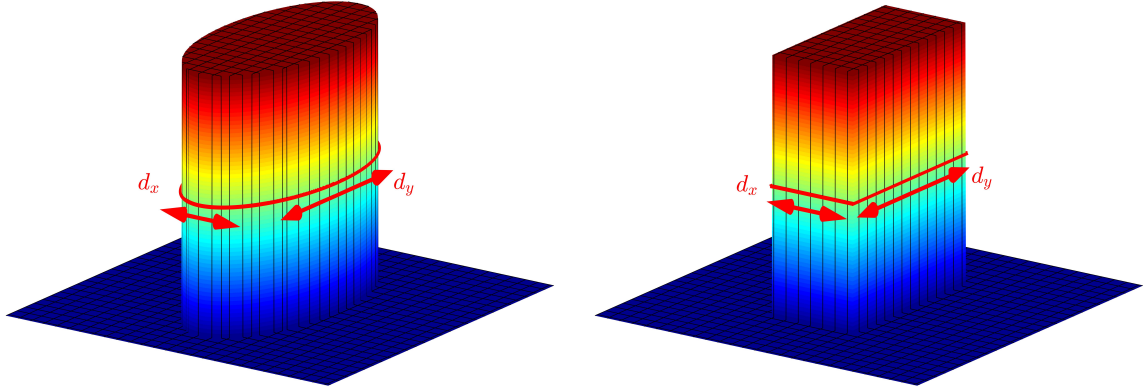

Figure 1. Depiction of an elliptical (left) and rectangular (right) tophat retinal irradiance profile used to simulate the thermal injury threshold value.

For applying the computer model which is validated against circular symmetric retinal images two issues have to be discussed. The first issue is the determination of the transmission which gives the amount of radiant energy that propagates through the retina. In case of circular symmetric spots, this was given by equation (3) where the effect of intraocular scattering is taken into account. Regarding non-uniform retinal spots, it is currently unclear how the transmission is determined. Therefore, we propose a new transmission equation which is based on equation (3). Here, equation (3) is integrated along the retinal spot shape and divided by the limits of the integration.

For an elliptical tophat irradiance profile with the widths $d_x$ and $d_y$ the averaged effective transmission is calculated by

$$T_{\text{eff}}^{\text{ell}}(\lambda, d_x, d_y) = \frac{2}{\pi} \int_0^{\frac{\pi}{2}} T_{\text{eff}}\left(\lambda, r^{\text{ell}}(\varphi)\right) \, d\varphi. \tag{7}$$

with

$$r^{\text{ell}}(\varphi) = \frac{d_x d_y}{4\sqrt{\left(\frac{d_x}{2}\sin(\varphi)\right)^2 + \left(\frac{d_y}{2}\cos(\varphi)\right)^2}}. \tag{8}$$

For a rectangular tophat irradiance profile the averaged transmission is calculated by

$$T_{\text{eff}}^{\text{rect}}(\lambda, d_x, d_y) = \frac{2}{\pi}\left(\int_0^{\varphi_{\text{C}}} T_{\text{eff}}\left(\lambda, \frac{d_x}{2\cos(\varphi)}\right) d\varphi + \int_{\varphi_{\text{C}}}^{\frac{\pi}{2}} T_{\text{eff}}\left(\lambda, \frac{d_y}{2\sin(\varphi)}\right) d\varphi\right) \tag{9}$$

with

$$\varphi_{\text{C}} = \arctan\left(\frac{d_y}{d_x}\right). \tag{10}$$

When it comes to small spot diameters which are in the region of the minimal spot size the transmission function strongly impacts the threshold value. To illustrate this behavior two further definitions of the transmission for non-uniform spots will be used in this investigation. In one definition the minimum of the spot diameters $d_x$ and $d_y$ will be used in equation (3). In the other the averaged spot size is inserted into equation (3).

The other issue is the definition of the minimal visible lesion (MVL) which is set as a circle with a diameter of $20\,\mu$m. However, in a non-uniform irradiance profile it is unclear how the shape of the MVL should look like. By using the definition of a circular symmetric MVL the temporal temperature behavior of the RPE layer in the direction of the smaller spot diameter has to be investigated. Using the temperature curve along the larger spot diameter leads to a higher temperature increase as there is less cooling and therefore leads to a lower threshold value. This issue dominates for spot shapes where one diameter is in the region of the minimal spot size and the other diameter is much larger. To illustrate the impact of the temperature curve, both the temperature in $x$- and $y$-direction of the MVL will be investigated.

## 5. REDUCTION FACTOR

The reduction factor (RF) is an essential value for laser and lamp safety issues.[3] All available emission limits for coherent and incoherent radiation[1,2,33,34] are derived from $ED_{50}$ values which represent the 50 % probability of injury and are understood as threshold values (THR). In this study, the threshold value is given in terms of the total intraocular energy. By applying the laser safety standard[2] to the investigated retinal images the corresponding maximum permissible exposures (MPE) are calculated. The reduction factor is then defined as

$$\text{RF} = \frac{\text{THR}}{\text{MPE}}. \tag{11}$$

The calculation of the MPE is performed by varying the angle of acceptance with a field stop in the retinal image.[7] This procedure is in accordance with the laser safety standard[2] and the ICNIRP guidelines.[1] This image analysis procedure is applied to the tophat profiles. Here, the diameters $d_x$ and $d_y$ are equal to the angular subtense of the apparent source $\alpha_x$ and $\alpha_y$ which is a term from the laser safety standard to describe the retinal spot size.[2] In Fig. 1, the field stops are illustrated by the solid red line representing the shape of the tophat.

The emission limits from the laser safety standard depend on the wavelength, the emission duration and the angular subtense $\alpha$ of the apparent source. The angular subtense is limited to the minimum spot size of 1.5 mrad and to the maximum spot size $\alpha_{\max}(t)$ which depends on the emission duration and varies from 5 mrad to 100 mrad. The average of both angular subtenses has to be calculated

$$\alpha = \frac{\alpha_x^{\lim} + \alpha_y^{\lim}}{2} \tag{12}$$

and inserted into the emission limit. The determination of the reduction factor is the main scope of this study as it directly shows the applicability of the image analysis in the thermal hazard region.

The definition of a suitable RF depends on several factors, e.g. the general uncertainty of the measured threshold data, the knowledge about the damage mechanism or the differences between the animal and the human. For this reason, the reduction factor does not always exactly reflect a safety margin between the threshold values and emission limit as there may be different uncertainties in the measured data. The greater the uncertainty is the greater the safety margin is set to derive the emission limit. Different damage mechanisms can lead to different RFs. For example, for the photochemical damage regime where the retina is irradiated for 100 s with a laser of the wavelength 441 nm the reduction factor is 23.[35] For the retinal thermal injury thresholds, a high RF is not needed as this injury mechanism is well understood and a RF of two is sufficient.[1]

## 6. RESULTS

### 6.1 Overview of simulation parameters

In the following, the computer model was applied to both tophat shapes from Fig. 1 for a wavelength of 530 nm as the RF is here the lowest for single pulses.[36,37] The simulation was performed on four parameter sets which are listed in Table 2. The intervals, depicted with the squared brackets, show a parameter variation. The retinal spot size is expressed by the angular subtense of the apparent source and can be calculated from the absolute size $d_x$ and $d_y$ on the retina. Here, the size of the irradiance profile in $x$-direction is set to the minimum spot size whereas in $y$-direction it is varied or set to 150 mrad which produces a long thin irradiated area on the retina. With these parameter sets the averaged spot size from equation (12) may show the highest discrepancy between the MPE and the THR.

Table 2. Parameter sets used for the computer model to predict the retinal thermal threshold values.

| parameter set | emission duration (s) | spot size $\alpha_x$ (mrad) | spot size $\alpha_y$ (mrad) |
|---|---|---|---|
| 1 | $[10^{-4}, 10]$ | 1.5 | 150 |
| 2 | $10^{-2}$ | 1.5 | [1.5, 150] |
| 3 | $10^{-1}$ | 1.5 | [1.5, 150] |
| 4 | 10 | 1.5 | [1.5, 150] |

## 6.2 Elliptical tophat spots

To illustrate the thermal behavior of the RPE layer, the isothermal curves for an elliptical tophat retinal irradiance are shown in Fig. 2 where the intraocular energy is set to the threshold value of about 375 µJ. Within the pulse duration of 1 ms the tissue is heated up inside the tophat profile and causes a strong temperature gradient at the edge of the elliptical profile. As a finite element method is used, the isothermal curves show fluctuations during the heating process. At the time 5.0 ms the hot spot is cooled down by the surrounding tissue and the temperature increase is less than half of the maximum temperature increase at the end of the pulse.

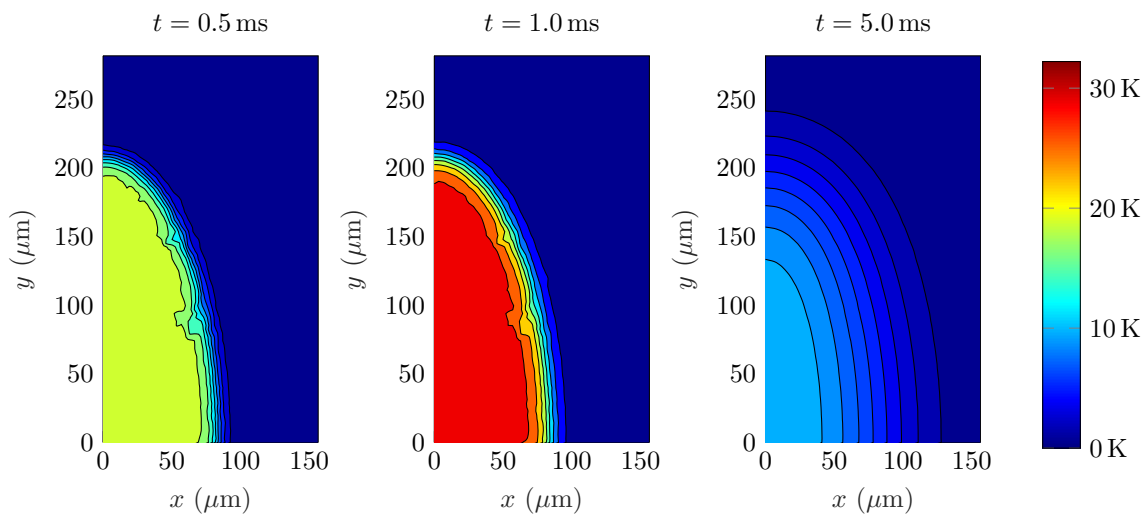

Figure 2. Isothermal curve for an elliptical tophat retinal irradiance profile with the widths $\alpha_x = 10$ mrad and $\alpha_y = 25$ mrad for a pulse duration of 1 ms. The isothermal curves show the temperature increase within the lesion depth of the RPE layer.

The results for parameter set 1 of Table 2 are shown in Fig. 3. The figure shows two essential results of the computer model which are depicted by the red and blue curves. The results depicted by the red curves are obtained by the temporal temperature behavior at the edge of the minimal visible lesion along the $x$-direction (THR$_x$) of the retinal irradiance profile whereas the blue curves show the results obtained by the temperature behavior along the $y$-direction (THR$_y$). Here, the corresponding threshold values were calculated by solving the relation $\Omega = 1$ with equation (1). According to the definition of a retinal thermal injury the minimal visible lesion is a circle with a diameter of 20 µm. With this definition the THR$_x$ value is seen as the correct result as this threshold value ensures to produce a damage within the MVL. Along the $x$-direction the temperature increases with a smaller slope at the edge of the MVL than along the $y$-direction as there is more cooling from the surrounding tissue. As a result, the $THR_x$ value is smaller than the $THR_y$ value. Both curves illustrate the dependence of the simulated threshold value from the definition of the MVL. In case of an MVL defined as an ellipse, the $THR_y$ could be considered as correct.

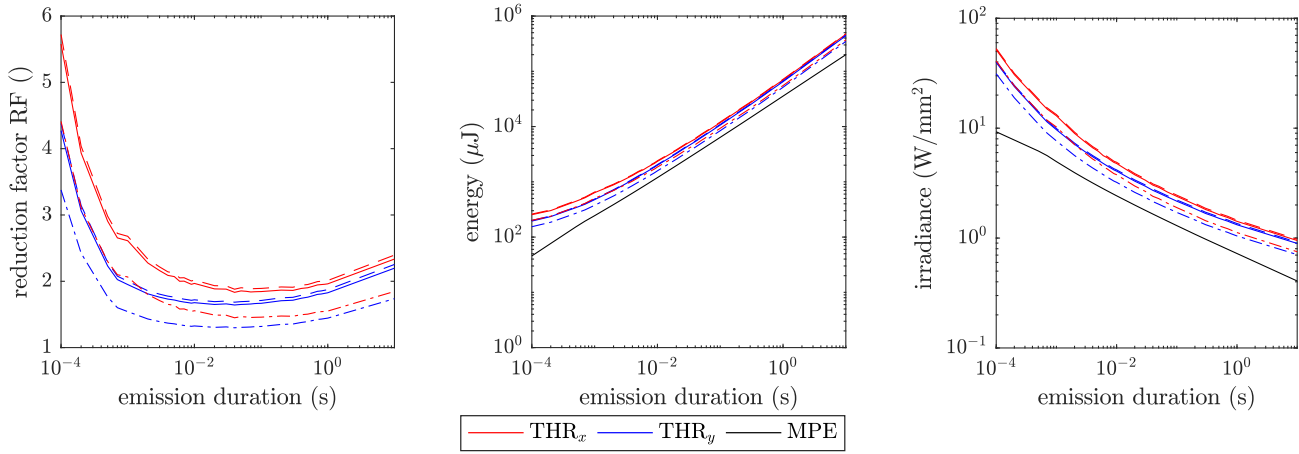

Figure 3. Reduction factor (left), energy (center) and irradiance (right) for the elliptical tophat and for parameter set 1 of Table 2 where the emission duration is varied. For the middle and right plot, the black solid line correspond the evaluation according to the laser safety standard. The red and blue curve show the threshold values according to the computer model for the consideration of the temperature behavior at the corresponding axis. The red and blue solid lines are based on the transmission function from equation (7). The corresponding dash-dotted line refers to a transmission calculation where the average spot size is inserted into equation (3) and the dashed line to a transmission where the minimum spot diameter is inserted into equation (3).

The solid red and blue curves show the results for our proposal according to equation (7) where non-uniform irradiance profiles are considered. The dashed curves are based on a transmission where the minimum spot diameter is inserted into the transmission function which can be applied for symmetric spot shapes. The transmission values are lower than in our proposed equation which leads to higher threshold values and therefore higher reduction factors. However, the differences are smaller than the comparison with the dashed-dotted curve where the averaged spot size is inserted into equation (3). Here, a higher radiant energy is assumed to propagate to the retina and therefore this consideration leads to more restrictive results.

For an emission duration of 40 ms, the reduction factor shows a minimum. For the solid red curve the reduction factor is 1.833 with a threshold value of 5.965 mJ. In case of the solid blue curve, the reduction factor is 1.642 with a threshold value of 5.344 mJ. The maximum permissible exposure according to the laser safety standard IEC 60825-1:2014[2] is 3.254 mJ. Even for a less restrictive consideration where the minimum spot diameter is used (dashed line) the minimum reduction factor for the red curve ($THR_x$) was 1.880 which is lower than the recommendation by the ICNIRP guidelines.[1]

The results for the parameter sets 2 to 4 of Table 2 are shown in Fig. 4. In these parameter sets the size of the irradiance profile along the $y$-direction was varied for a constant emission duration. The impact of the transmission behavior is higher for an increasing retinal spot size. In case of the first parameter step $\alpha_y = 1.5$ mrad all transmission functions show the same value which leads to the same threshold and reduction factor values. Similar to the results of parameter set 1 from Fig. 3, the difference between our proposal from equation (7) (solid line) and using the minimum spot diameter for equation (3) (dashed line) is smaller than using the average spot size (dash-dotted line).

The right plots of Fig. 4 show the irradiance for the threshold values obtained by the computer model and the emission limits expressed by the irradiance according to the laser safety standard IEC 60825-1:2014.[2] Here, the irradiance converges to a constant value which is caused by the maximum spot size limitation from the laser safety standard and the creation of a local hot spot in the RPE layer where the cooling of the surrounding tissue is not sufficient to allow a higher irradiance level.

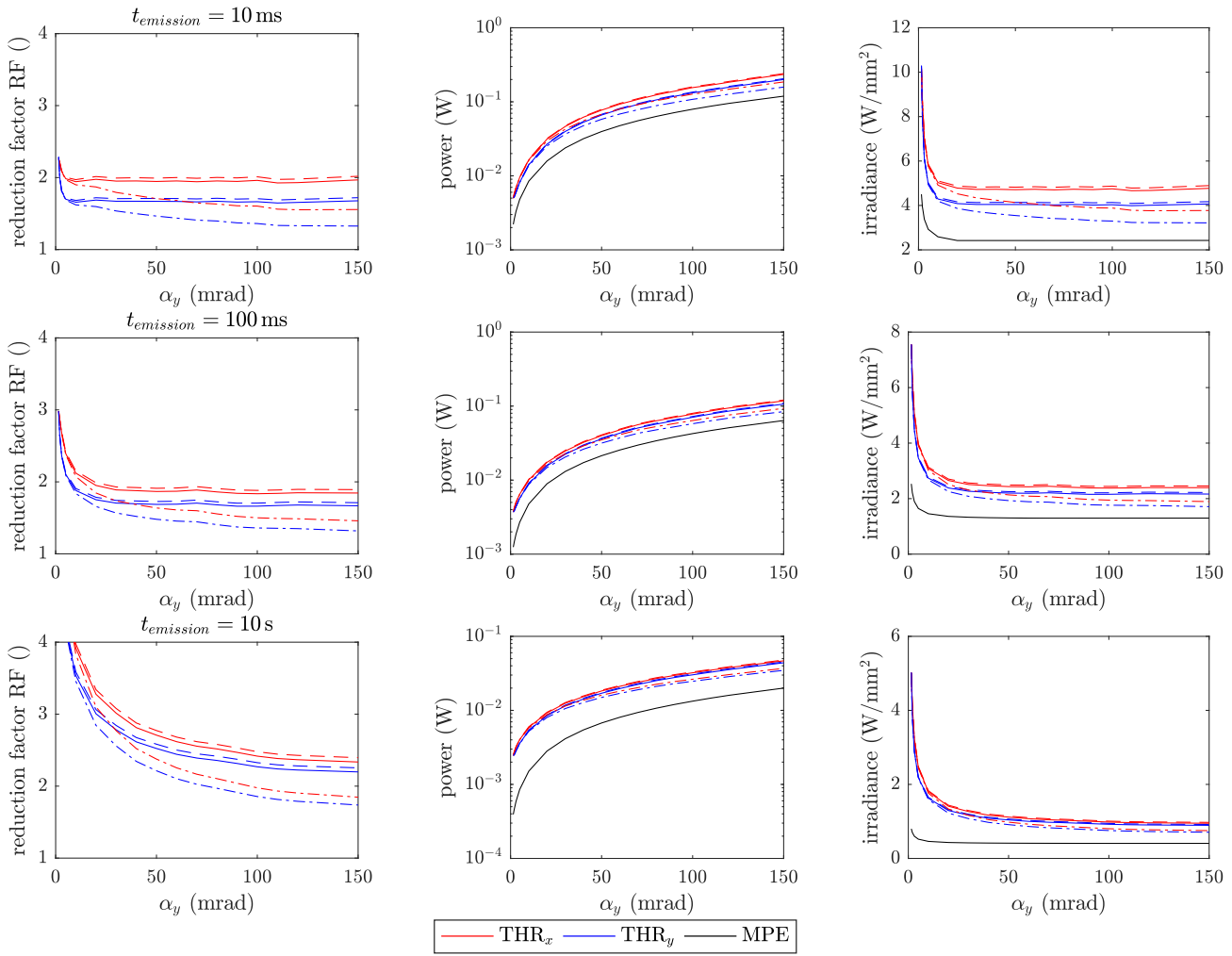

Figure 4. Reduction factor (left), power (center) and irradiance (right) for the elliptical tophat and for parameter sets from 2 to 4 of Table 2 where the retinal spot size is varied in $y$-direction. Further details are the same as in Fig. 3.

By considering the transmission behavior from equation (7) the following minimum reduction factors listed in Table 3 can be found. The minimum values for the $x$- and $y$-direction do not always appear for the same size $\alpha_y$ which is the case for parameter set 3. The minimum reduction factor regarding the $\text{THR}_x$ value is about 1.8 and regarding the $\text{THR}_y$ about 1.6. If the different transmission functions are taken into account the minimum values are 1.892 (dashed line) and 1.459 (dash-dotted line) for the $\text{THR}_x$ value and 1.720 (dashed line) and 1.326 (dash-dotted line) for the $\text{THR}_y$ value regarding the same parameter set.

Table 3. Minimum reduction factors for the parameter set from 2 to 4 (Table 2, Fig. 4).

| parameter set | $x$-direction | | | | $y$-direction | | | |
|---|---|---|---|---|---|---|---|---|
| | RF () | $\alpha_y$ (mrad) | $\text{THR}_x$ ($\mu$J) | MPE ($\mu$J) | RF () | $\alpha_y$ (mrad) | $\text{THR}_y$ ($\mu$J) | MPE ($\mu$J) |
| 2 | 1.926 | 110 | 1.682e3 | 0.873e3 | 1.647 | 110 | 1.438e3 | 0.873e3 |
| 3 | 1.835 | 100 | 7.799e3 | 4.250e3 | 1.662 | 90 | 6.355e3 | 3.824e3 |
| 4 | 2.335 | 150 | 467.019e3 | 199.980e3 | 2.198 | 150 | 439.469e3 | 199.980e3 |

## 6.3 Rectangular spots

In comparison with the elliptical tophat, the reduction factors are larger for the rectangular tophat profiles. The minimum reduction factor also appears for an emission duration of 40 ms. For the solid red curve the reduction factor is 2.332 with a threshold value of 7.588 mJ whereas for the solid blue curve the reduction factor is 2.089 with a threshold value of 6.798 mJ. According to the laser safety standard, the maximum permissible exposure is 3.253 mJ. Furthermore, the irradiance which represents both the threshold and the maximum permissible exposure is lower in case of the rectangular tophat than for the elliptical tophat. This is due to the thermal behavior where the cooling is better for the elliptical shape. In case of the rectangular tophat there is less heat dissipation in the center point of the irradiance profile.

For all parameter sets from 1 to 4 the maximum permissible exposure according to the laser safety standard IEC 60825-1:2014[2] is the same as it was for the elliptical distribution. This is due to the fact that different field stop shapes were used for the laser safety evaluation.

By considering the transmission behavior from equation (7), the following minimum reduction factors listed in Table 4 can be found. The minimum reduction factor regarding the $x$-direction is about 2.3 and regarding the $y$-direction about 1.9. By taking the different transmission functions into account, the minimum values are 2.330 (dashed line) and 2.293 (dash-dotted line) for the $\text{THR}_x$ value and 1.993 (dashed line) and 1.962 (dash-dotted line) for the $\text{THR}_y$ value regarding the same parameter set.

Table 4. Minimum reduction factors for the parameter set from 2 to 4 (Table 2).

| parameter set | $x$-direction | | | | $y$-direction | | | |
|---|---|---|---|---|---|---|---|---|
| | RF () | $\alpha_y$ (mrad) | $\text{THR}_x$ ($\mu$J) | MPE ($\mu$J) | RF () | $\alpha_y$ (mrad) | $\text{THR}_y$ ($\mu$J) | MPE ($\mu$J) |
| 2 | 2.303 | 5.0 | 110.468 | 47.961 | 1.970 | 5.0 | 94.502 | 47.961 |
| 3 | 2.317 | 30 | 3.028e3 | 1.307e3 | 2.098 | 30 | 2.742e3 | 1.307e3 |
| 4 | 2.880 | 120 | 460.499e3 | 159.876e3 | 2.714 | 120 | 433.854e3 | 159.876e3 |

## 7. SUMMARY AND CONCLUSION

A three dimensional computer model to simulate the retinal thermal injury damage mechanism by solving the heat equation and the Arrhenius equation with the temporal temperature behavior was used to predict the threshold value of laser-induced retinal injuries for non-uniform irradiance profiles. The computer model was derived from the work of Jean[13,14] whose model is validated against experimental $ED_{50}$ values of non-human primates. We investigated an elliptical and a rectangular tophat as retinal irradiance distribution and compared the simulated threshold values with the corresponding emission limits according to the laser safety standard IEC 60825-1:2014.[2] Currently no measured threshold values exist for these profiles since experiments with non-uniform profiles have not been performed in the past.

We examined the applicability of the MVL definition to non-uniform retinal irradiance profiles whereas two different threshold values were simulated. Both threshold values are obtained with $\Omega = 1$ for the two orthogonal directions along the retinal spot shape and show how the predicted threshold value changes. Here, four different parameter sets were investigated. In one parameter set the emission duration was varied where in the other sets the retinal spot size was varied in one direction which shows the transition from the minimum spot size to an elongated spot. According to the laser safety standard the averaged spot size has to be considered. The reduction factor which is calculated by the ratio between the threshold value and the maximum permissible exposure according to the laser safety standard directly shows the validity of this calculation procedure. In this study, the minimum reduction factor is 1.926 for the MVL definition of the smaller extend and 1.662 for the other direction, respectively. These minimum values are below the reduction factor of two which is defined by the ICNIRP guidelines to be sufficient.[1] According to the definition of a circular MVL, the minimum reduction factor is 1.926 for the investigated non-uniform irradiance profiles. This is sufficient and proofs the applicability of the calculation procedure from the laser safety standard in these cases.

In addition, different transmission behaviors were investigated. For small spot sizes intraocular scattering dominates where the direct transmission has to be taken into account. In case of symmetric spot shapes, Jean[13,14] proposed an equation that ensures a transition from direct to total transmission. On this base, we propose a method which can be applied to non-uniform irradiance profiles. To illustrate the influence of the transmission on the threshold value, two different methods were additionally investigated. In one method the minimum spot diameter was used and in the other method the averaged spot size was used in the formula for symmetric spots. In our results, the method with the minimum spot size is less restrictive and leads to reduction factors up to 2.6 % higher than in our proposed method. The transmission based on the average spot size results in more restrictive evaluations which are up to 21 % smaller than in our method as less intraocular scattering is taken into account. Therefore, this method is seen as inappropriate.

At last, the investigation of elliptical shaped and rectangular shaped tophat profiles shows that the threshold values expressed by the total introcular energy is higher for rectangular tophats. However, with regard to the corresponding irradiance the rectangular tophat shows lower values than the elliptical tophat. This behavior can be explained by different heat dissipations caused by the surrounding tissue. In case of the rectangular tophat the cooling is weaker than for elliptical tophats.

## 8. OUTLOOK

In a future study, this issue of the MVL can be avoided by setting the minimum spot diameter of the irradiance profile to 5 mrad which is about four times larger than the MVL. In addition, using a larger minimum spot diameter avoids several other issues that come with the minimum spot size in laser safety evaluations.[31]

The other issue is the transmission behavior for non-uniform retinal images. With a similar approach as described above where the minimum spot diameter is set to 5 mrad this problem can be reduced but not solved. To get the transmission behavior for retinal images with widths of the minimum spot size the exact scattering behavior has to be taken into account. In this study a procedure was developed to calculate the transmission for a non-uniform retinal spot. However, this transmission factor is treated as a constant factor that is multiplied to the irradiance profile on the retina. As a consequence, the reduction factor scales linearly with the transmission factor. In the region of the minimum spot size, this approach is currently only validated for radially symmetric retinal images.

Also, in a future study an elliptical Gaussian distribution can be investigated with regard to the simulated threshold value and the maximum permissible exposure. In addition, as there is currently no explicit definition,[38] an elliptical as well as a rectangular field stop can be used for the laser safety evaluation.[2] Here, the differences between both approaches can be discussed and compared to the simulated threshold values.

At last, the same study can be expanded to the topic of pulsed irradiance profiles. The calculation procedure of the laser safety standard IEC 60825-1:2014[2] for pulsed retinal images shows a more complex evaluation where three criterions have to be applied. Here, the applicability of the laser safety evaluation method can be verified for non-uniform irradiance profiles.

## 9. ACKNOWLEDGEMENTS

The authors would like to thank Mathieu Jean and Dr. Karl Schulmeister from the Seibersdorf Labor GmbH for inspiring discussions regarding the simulation of retinal thermal injury thresholds.